# Non-equilibrium States and Interactions in the Topological Insulator and Topological Crystalline Insulator Phases of NaCd$_4$As$_3$

Tika R Kafle[1], Yingchao Zhang[1], Yi-yan Wang[2], Xun Shi[1], Na Li[1], Richa Sapkota[1], Jeremy Thurston[1], Wenjing You[1], Shunye Gao[1], Qingxin Dong[2], Kai Rossnagel[3,4], Gen-Fu Chen[2], James K Freericks[5], Henry C Kapteyn[1,6], Margaret M Murnane[1]

[1]Department of Physics and JILA, University of Colorado and NIST, Boulder, CO 80309, USA
[2]Institute of Physics and Beijing National Laboratory for Condensed Matter Physics
[3]Institute of Experimental and Applied Physics, Kiel University, D-24098 Kiel, Germany
[4]Ruprecht Haensel Laboratory, Deutsches Elektronen-Synchrotron DESY, D-22607 Hamburg, Germany
[5]Department of Physics, Georgetown University, Washington, DC 20057, USA
[6]KMLabs Inc., 4775 Walnut Street, #102, Boulder, Colorado 80301, USA

* Corresponding author: tika.kafle@colorado.edu

**Abstract**

Topological materials are of great interest because they can support metallic edge or surface states that are robust against perturbations, with the potential for technological applications. Here we experimentally explore the light-induced non-equilibrium properties of two distinct topological phases in NaCd$_4$As$_3$: a topological crystalline insulator (TCI) phase and a topological insulator (TI) phase. This material has surface states that are protected by mirror symmetry in the TCI phase at room temperature, while it undergoes a structural phase transition to a TI phase below 200 K. After exciting the TI phase by an ultrafast laser pulse, we observe a leading band edge shift of >150 meV, that slowly builds up and reaches a maximum after ~0.6 ps, and that persists for ~8 ps. The slow rise time of the excited electron population and electron temperature suggests that the electronic and structural orders are strongly coupled in this TI phase. It also suggests that the directly excited electronic states and the probed electronic states are weakly coupled. Both couplings are likely due to a partial relaxation of the lattice distortion, which is known to be associated with the TI phase. In contrast, no distinct excited state is observed in the TCI phase immediately or after photoexcitation, which we attribute to the low density of states and phase space available near the Fermi level. Our results show how ultrafast laser excitation can reveal the distinct excited states and interactions in phase-rich topological materials.

**Introduction**

Topological materials can host unique conducting surface and edge states that are robust against disorder, defects, and modification. In addition, they exhibit interesting surface properties such as spin polarization, momentum locked surface spin[1-2], and backscattering suppression,[3] which makes them promising platforms for technological applications. Due to these unique properties, the search for candidate topological materials for novel technological applications has significantly increased in the past 15 years[4]. To date, many 2D and 3D topological insulators (TIs), including semi-metallic TIs, dual topology insulators,[5-6] and topological crystalline insulators (TCIs) have been theoretically predicted[4] – however, only a few of them have been experimentally realized.

TCIs are a new class of quantum materials in which the surface band topology is protected by crystal space group symmetries[7-10], unlike time reversal symmetry in a non-trivial Z2 topological insulator (TI)[8-9, 11-12]. In general, these materials possess a bulk band gap with the coexistence of bulk-boundary gapless Dirac-cone-like surface states. Owing to the rich variety of crystalline symmetries (230 crystalline space groups)[13], many TCIs have been identified with surface states protected by mirror symmetries, glide mirror symmetries and more recently rotational protected symmetries[14]. In addition, these materials undergo topological phase transitions by tuning the composition, concentration, temperature[15-16], and pressure[17-19]. Being phase rich materials, their non-equilibrium states might exhibit many-body physics related to electrons, spins, phonons, polarization, bulk-surface carrier dynamics, and carrier relaxation channels – which have been probed by photoemission[29-32] and optical techniques[20-22]. However, only a few studies to date have explored the non-equilibrium states and interactions in these materials, which can also reveal novel information about the ground state interactions.

Time- and angle-resolved photoemission spectroscopy (tr-ARPES) is a powerful method for probing the dynamic electronic order of quantum materials with energy and momentum resolution. This makes it possible to track surface and bulk carrier dynamics, as well as the coherent many-body interactions between charges, phonons, and spins.[23-25] Most studies to date have probed Z2 TI materials, and in particular, group IV-VI compounds and their derivatives involving alloys that lead to a TCI or TI, as well as prototype TI compounds containing Bi.

Here we study the ultrafast response and relaxation of $NaCd_4As_3$, a material that exhibits dual topology – a topological crystalline insulator phase at room temperature and a topological insulator phase at temperatures below ~190 K. In the low temperature TI phase, although only weak photoexcitation is observed during the ~40 fs laser pulse, a very large chemical potential shift of >150 meV within $\Gamma \pm 0.15$ Å$^{-1}$ near the Fermi level ($E_F$) is observed, which maximizes after ~0.6 ps and persists for ~8 ps. Moreover, the electron temperature ($T_e$) also maximizes on similar timescales of ~0.6 ps, which is long after the laser excitation pulse. This behavior and its timescales suggest that the electronic and lattice orders are strongly coupled and that the induced band structure changes are due to partial relaxation of the lattice distortion that is known to be associated with the TI phase (see Fig. 1).[26] Furthermore, the delayed rise time of electron temperature suggests that the directly excited electronic states and the electronic states near $E_F$ probed by tr-ARPES are weakly coupled, possibly also due to the partial relaxation of the lattice distortion. The long persistence time of ~8 ps for the chemical potential ($\mu_F$) shift, excited electron distribution and enhanced electron temperature are likely related to heat transport from the surface to the bulk, before the ground state TI distortion recovers. In contrast, in the room temperature TCI phase, we observe a very weak excited carrier distribution and minimal chemical potential shift of <30 meV after laser excitation. Our findings are among the first to probe non-equilibrium dynamics and interactions in both the TCI and TI phases of a topological material in its pure form.

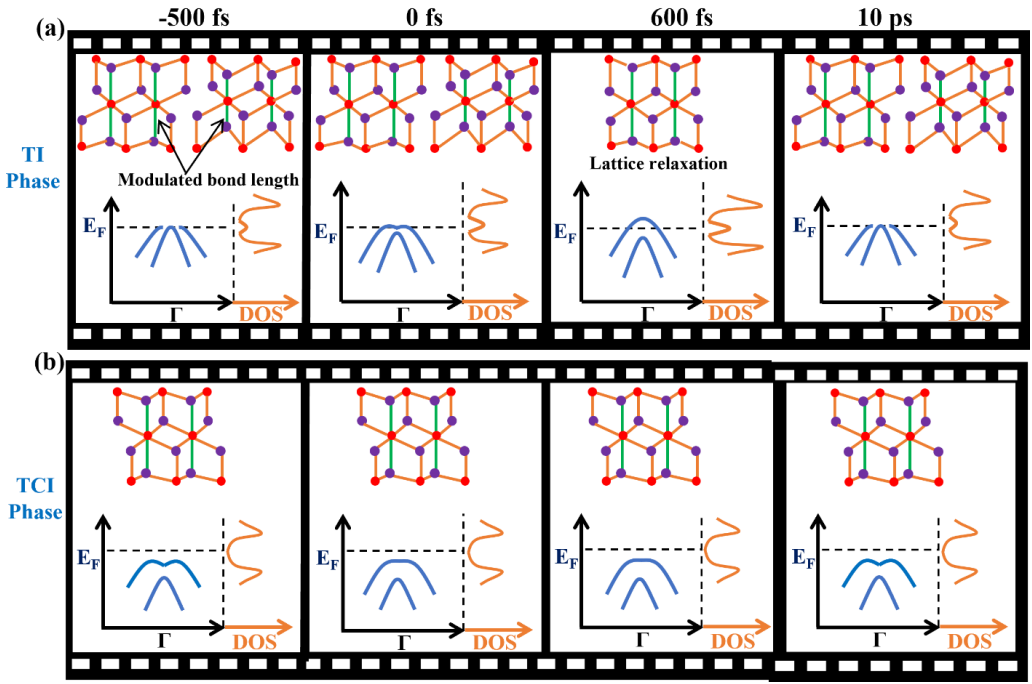

**Figure 1:** Schematic of the coupled evolution of the lattice and electronic orders and DOS near the Fermi level of (a) TI phase and (b) TCI phase after excitation by an ultrafast laser pulse. The electronic band and DOS plotted above is intended to guide the reader to reflect the change induced by the pump pulse, but should not be read as an exact band dispersion and DOS. The lattice structure shown is a section of the Cd-As substructure (filled circles: purple – Cd and red – As atom). The Cd-As bond length (green line indicated by arrowhead) is modulated in the TI phase, but remains unchanged in the TCI phase.[27] For a schematic of the full TCI phase, see S1.

## Results

To study the excited state dynamics of NaCd$_4$As$_3$ in both the TCI and TI phase, a laser pump pulse (1.58 eV photon energy, 40 fs pulse duration) was used to excite the material. The band structure dynamics were then tracked using a 22.1 eV, 10 fs duration, probe pulse, and the resulting photoemitted electrons are analyzed using an angle-resolved photoemission spectroscopy setup (see materials and methods section for pump fluence and other details). NaCd$_4$As$_3$ is an n-type semi-metallic[26, 28] TCI at temperatures above ~190 K. Below this temperature, it transitions from a monoclinic Cm to a rhombohedral R$\bar{3}$m space group, corresponding to a topological phase transition from a TCI to a TI phase. In the TCI phase, the surface states are protected by the mirror symmetry of the (110) plane at the Γ and T points (see S1). Below the transition temperature, the mirror symmetry is broken; however, the band inversion at the Γ point persists, as it is protected by time-reversal symmetry. The observed phase transition in our ARPES data before the pump pulse agrees well with previous static ARPES reports[28] (see S1), within our energy resolution.

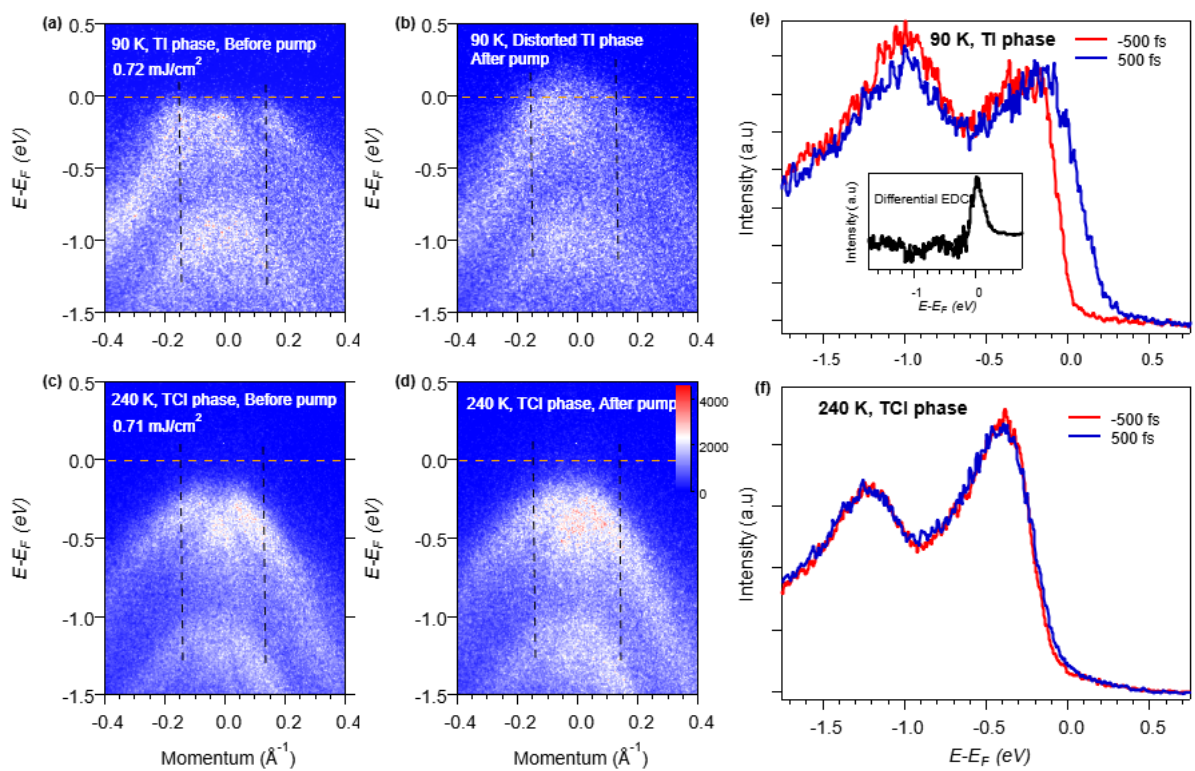

**Figure 2:** 2D ARPES plots of NaCd$_4$As$_3$ taken at 90 K (a, b) and 240 K (c, d); (a, c) before and (b, d) after laser excitation. The horizontal dotted brown line indicates the Fermi level. The corresponding energy distribution curves (EDCs), integrated over the momentum range indicated by the vertical dashed line, at 90 K and 240 K are shown in (e) and (f), respectively. The inset in (e) is a differential EDC curve before and after laser excitation.

Typically in a topological material, bulk bands and surface states (SS) coexist near $E_F$ – in which case electrons would initially be excited from both bulk bands and surface states[24] into unoccupied states. In the TCI phase as shown in Fig. 2, few excited state carriers are observed after the pump pulse. In contrast, in the low-temperature TI phase, as shown in Figs. 2 and 3, the 40-fs laser pulse initially excites a small population of carriers. Then, a transient state starts to evolve and maximizes on timescales of ~0.6 ps (Fig. 3), which is far slower than the duration of the laser excitation pulse. The transient state appears to be continuous across $E_F$, i.e., states above and below $E_F$ are populated. To determine the leading band edge shift, energy distribution curves (EDCs), integrated at ~ $\Gamma \pm 0.15$ Å$^{-1}$ (black dotted vertical lines in Fig. 2), were plotted (Figs. 2e, f), and Fermi-Dirac fit and Fermi-edge center methods were applied (see S2 for the energy shift determination technique). The results of these two methods are in good agreement. In particular, in the line-shape analysis of the TI phase data, we find that only varying the position and width of the Fermi edge can consistently reproduce the observed dynamics, while varying the position or width of the leading peak gives inconsistent fit results (see S2). The stability of peak position and width can be read directly from the TCI phase data (Fig. 2f). In the TI phase, an upward energy shift of the leading edge of ~170 meV was obtained, but only ~30 meV in the TCI phase. We note that $\Delta\mu_F$ in the TI phase can vary by ±25 meV depending on the choice of integration interval of the momentum-integrated EDCs (see S3). We refer to the leading edge shift resulting from the filling of the unoccupied states as the chemical potential shift ($\Delta\mu_F$). This chemical potential shift exhibits a long persistence time of ~8 ps, followed by a relaxation of the excited electron distribution and electron temperature. In addition, the chemical potential shift is also accompanied by an increase in the electron temperature from the equilibrium value of 90 K to ~800 K at the peak value of $\Delta\mu_F$ (see Fig. 3d).

    We further note that the delayed rise time of the electron temperature, $T_e$, (Fig. 3d) is in sharp contrast to the very fast rise time that is observed in most materials probed by tr-ARPES, which peaks during the laser excitation pulse. For the TI phase of NaCd$_4$As$_3$, the ARPES spectra show no significant excited state population immediately after laser excitation. Hence, the delayed increase of $T_e$ suggests that the Cd 4$d$ and As 4$p$ orbitals we probe (near E$_F$) are different from the orbitals occupied by the hot electrons immediately after excitation. However, after partial relaxation of the lattice distortion in ~0.6 ps – leading to changes in band structure and orbital

occupancy – the ARPES spectra are sensitive to these hot electrons, as shown in Fig. 3(d). They are also sensitive to the polarization of the EUV probe.

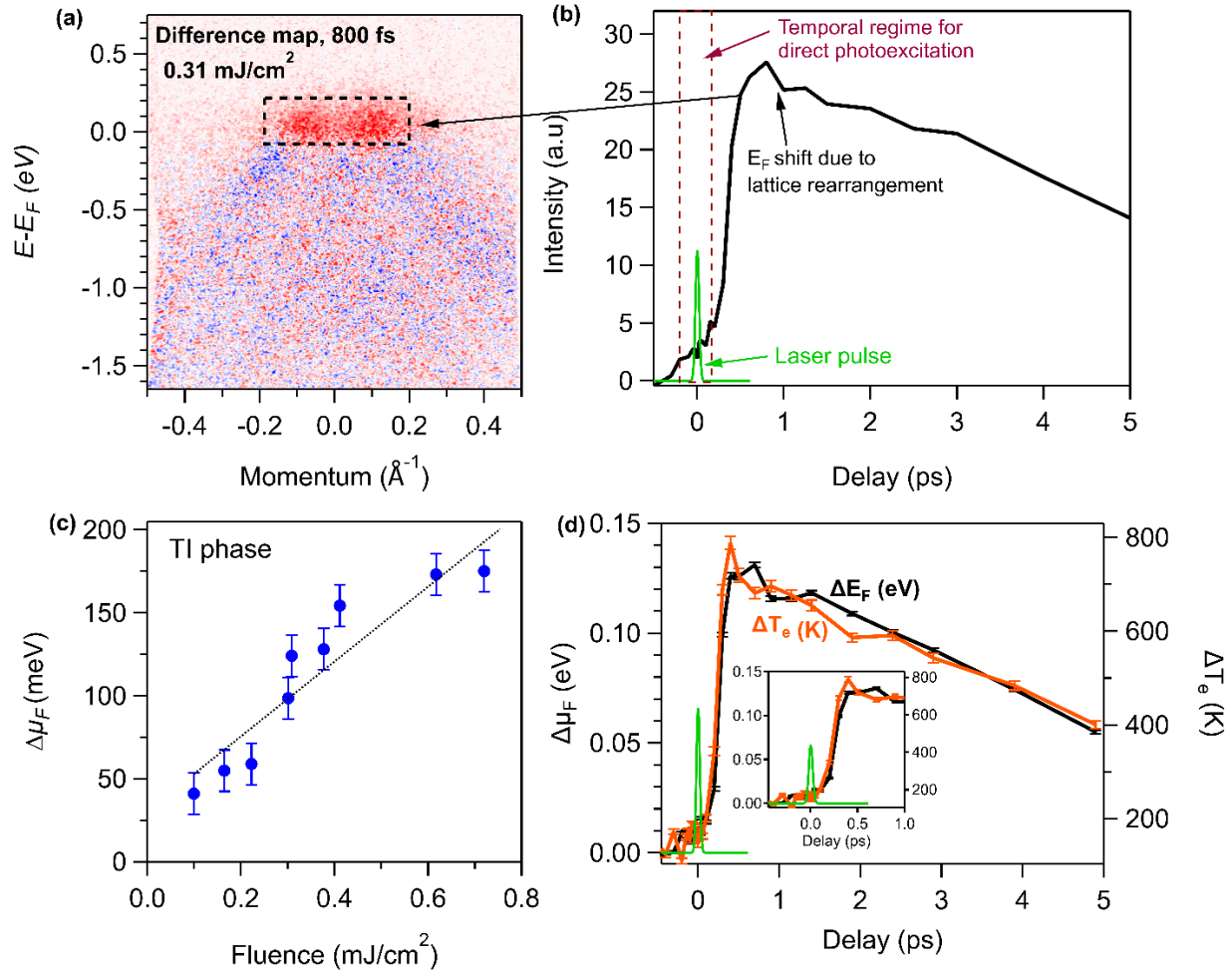

**Figure 3**: (a) Differential tr-ARPES map before and after laser excitation (800 fs) of the TI phase at 90 K. (b) Electron dynamics integrated over the momentum and energy range indicated by the black dotted box in (a). (c) The maximum change in chemical potential shift as a function of pump laser fluence. The error bar indicates the $\Delta\mu_F$ range obtained by varying momentum integration area centered at $\Gamma$ to obtain EDCs of interest for data analysis (see S3). (d) Change in electron temperature ($\Delta T_e$) and chemical potential shift ($\Delta\mu_F$), both obtained from the Fermi-Dirac distribution.

To better understand the chemical potential shift in the low-temperature TI phase, we investigated the electronic band structure dynamics, which are plotted in Fig. 3. In Fig. 3a, we integrated the electron population (black dotted box) above the ground state TI Fermi level and plotted the dynamics in Figs. 3b and d. The chemical potential shift builds up slowly and the peak occurs well after the laser pump pulse (see inset of Fig. 3d). This suggests that the changes we observe may be due to a relaxation of the TI lattice deformation, which modulates both the atomic lattice and electronic band structure, in agreement with a recent report.[29] A schematic of the predicted lattice bond length modulation[27], and the observed dynamic electronic structure and density of states after laser excitation is shown in Figure 1. The largest bond length modulation of nearly 0.7 Å is predicted to occur for the Cd-As bond (see first panel of Fig. 1a, green line) in the ground state of the low-temperature TI phase[27], which are the bands we probe with tr-ARPES. Such a modulation is not observed in the TCI phase.[27] Furthermore, the excited carriers appear around Γ along the K-Γ-K direction – a full mapping in the Γ-K direction shows no clear evidence of direct photoexcitation away from the Γ-point. This rules out the explicit involvement of charge transfer processes.

To better understand the coupling of the charge and lattice orders, we fit the band edge to a Fermi-Dirac distribution (see S2) as a function of time delay to extract the change in both the electron temperature ($\Delta T_e$) and the chemical potential shift ($\Delta \mu_F$) (see Fig. 3d).[30-31] We observe the similar slow rise time (~600 fs) of both the chemical potential shift and the electron temperature – once again on timescales far longer than the 40 fs laser excitation pulse. Processes such as lattice distortion relaxation or electron scattering from higher excited states could in theory account for the observed slow rise time.[30, 32] For example, a similar timescale of ~700 fs has been observed in the TI material $Bi_2Se_3$ for excited electrons to scatter from higher lying states to lower bulk states and/or surface states, followed by a ~1.7 ps timescale for the electron energy relaxation channel via electron-phonon coupling.[30, 32] However, since higher excited states are not observed in our data, this cannot explain the observed slow rise time.

Next, we scanned the pump laser fluence to determine if there was evidence of coherent phonon and/or selective phonon mode excitation in the 2D surface states of $NaCd_4As_3$. The chemical potential shift shows a linear dependence on the pump laser fluence, as shown in Fig. 3c. The maximum $\Delta \mu_F$ of ~170 meV was observed for a laser fluence of 0.72 mJ/cm$^2$, which is the maximum fluence that we can use while still avoiding space charge distortion effects. We note that for some data sets (different sample pieces), $\Delta \mu_F$ ~170 meV could be obtained with a slightly lower pump fluence, but further increase of the pump fluence caused spectral distortion. In contrast, at similar laser fluence, a very small chemical potential shift was observed in the TCI phase (see Fig. 2f and S3). We found a gradual rise in $T_e$ with pump fluence – however, the rise time is independent of the laser fluence (see S4), further supporting strong electronic and lattice order coupling. The excited state exhibits a single exponential decay with a time constant of 4.25 ps, which is consistent with the timescale observed in most TI materials.[15, 25, 33-35] We note that because of the narrow bulk band gap of semi-metallic NaCd$_4$As$_3$, bulk band and surface states coexist near E$_F$, for both equilibrium and out-of-equilibrium states. Although we do not resolve these states, the dynamics are dominated by the lattice distortion and not by the lifetimes of the excited bulk or surface electronic states. Finally, the long relaxation time of ~8-10 ps for the material to return to the equilibrium TI state (Fig. 4) indicates that the relaxation mechanism in NaCd$_4$As$_3$ is likely due to electron-phonon and phonon-phonon relaxation, and heat transport from the surface to the bulk.

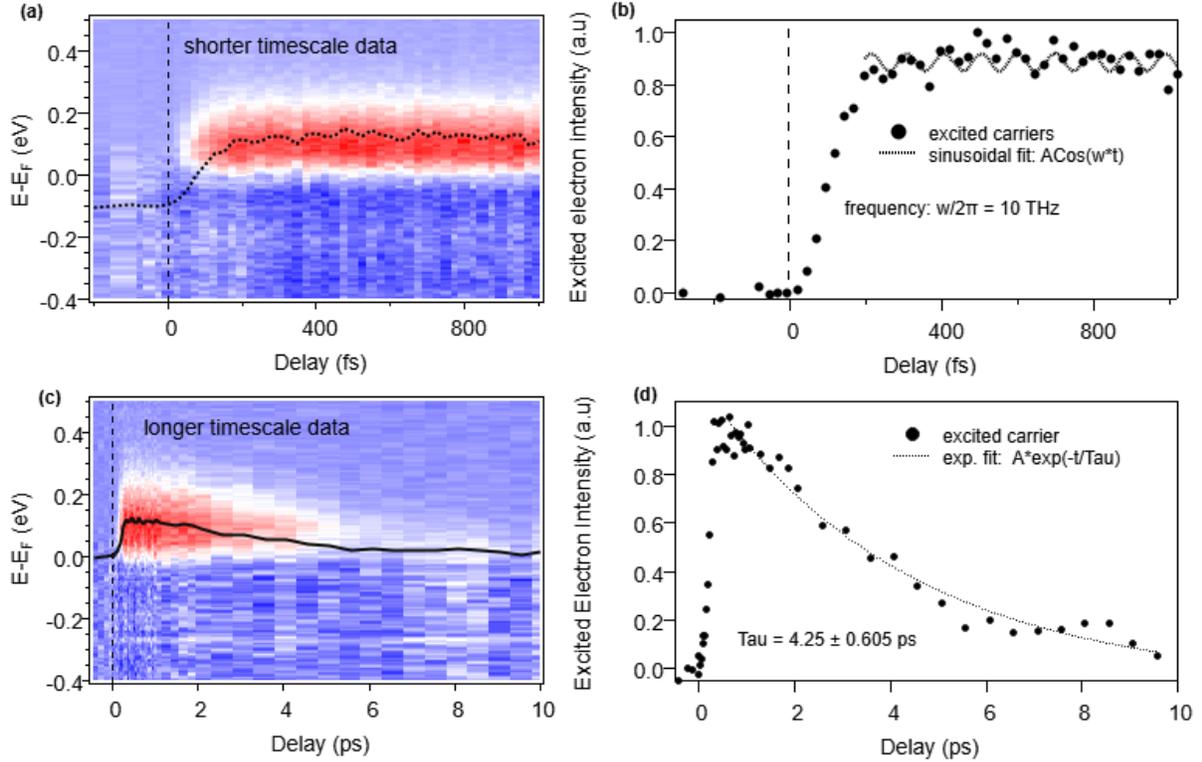

**Figure 4**: 2D temporal map integrated within Γ ± 0.15 Å of a differential ARPES spectrum measured at 78 K for (a) a shorter timescale and (c) a longer timescale. The black lines are the 1D temporal dynamics of the same excited electron carriers which are plotted in (b) and (d) respectively for better visualization. The dashed lines in (b) and (d) are the sinusoidal and exponential fits, respectively.

Some weak oscillations (Fig. 4) were observed in the excited electron intensity within the first few ps, and a period of 100 fs (frequency of 10 THz) was obtained from the sinusoidal fit. Such weak oscillatory signatures are also observed in the Δμ$_F$ and ΔT$_e$ obtained from a Fermi-Dirac fit (see S2 and S5). We speculate that the origin of these oscillations is associated with charge delocalization after excitation of the material, which can excite strongly coupled phonon modes and partially relax the lattice distortion in the TI phase (i.e., the Cd-As bond length modulation).

To better resolve the phonon mode frequency, measurements were taken at different laser fluences – a Fast Fourier Transform (FFT) indicated a frequency in the 10–14 THz range (see S5). Unfortunately, due to the constraints of surface degradation of the material over time (see Method and Material section) and the weak nature of the oscillation amplitude itself, it was not possible to take data over a long period to improve the signal-to-noise ratio - hence limiting the determination of the precise frequency of the phonon modes involved. We note that a ~5.7 THz Raman phonon mode in NaCd$_4$As$_3$ has been reported for its room temperature TCI phase.[27] The observed increase in phonon frequency modes in the TI phase could result from either the relaxation of the distorted TI lattice, leading to a shortening of the bond length and an increase in the frequency. Or, if the pump laser pulse excites charges into anti-bonding orbitals of Cd-As atoms, it doubles its intrinsic frequency mode.

**Discussion:**

There are studies reporting that even small lattice distortions or displacements can induce a topological phase transition.[15, 18] Our data, combined with other recent work[29], suggest that strong coupling of the electronic and structural (lattice) orders is responsible for the observed slow rise time of the transient state population, the chemical potential shift, and the increase in electron temperature ($\Delta T_e$), which closely track each other. Such strong couplings have recently been observed in 2D TaSe$_2$, where optical excitation excites a high amplitude breathing mode, which in turn modulates the electron temperature, in an isolated isentropic system.[30] We note that the large time lag (~0.6 ps) observed for the electron temperature to peak is in contrast with most materials studied to date using ultrafast laser excitation. In most systems, the laser energy is rapidly absorbed by the electrons, so that their temperature peaks within (or soon after) the laser pulse. The very large time lag observed in NaCd$_4$As$_3$ could arise if electrons from either or both Cd 4$d$ and As 4$p$ orbitals are initially excited by the laser into delocalized states/orbitals with a symmetry that is not accessed by the EUV probe pulse. Only after the partial relaxation of the lattice distortion and the associated band structure changes, can the hot electrons be probed. Indeed, such sensitivity of orbital bands and their associated odd/even (parity) states to light polarization in photoemission has already been observed.[36-38] Previous work has shown that lattice distortion of the TI phase can relax at only slightly elevated temperatures of $\Delta T$ ~10 K (far less than what is required to drive an equilibrium TI to TCI phase transition).[39] The electronic order will follow the evolving structural order, with the rise and fall times for the electron temperature and chemical potential dictated by the lattice distortion relaxation and recovery, as well as cooling via heat transport into the bulk.

Specifically, the transient excited state that lasts for ~8 ps and that is observed only in the excited TI phase could arise from relaxation of the lattice distortion in the a-b plane of $NaCd_4As_3$ associated with the TI phase. In particular, the atomic bond length between Cd-As along the *c*-axis in the $Cd_6As$ octahedra within the $^{\infty}_{2}[Cd_4P_3]^{-}$ (a 2D polyanionic substructure) has a significant modulation at low temperature – by ± 0.3 Å (see Fig 1a, first panel) at 130 K.[26-27] This bond length modulation shifts the relative distance between the $Na^+$ and $^{\infty}_{2}[Cd_4P_3]^{-}$ sublattices, causing charge redistribution and hence electronic band structure changes (Fig. 1a, c).[15] The density of states (DOS), contributed mainly by Cd and As atoms near $E_F$, increases – providing an increased phase space for excited carriers. Furthermore, the laser-induced reduction in the Cd-As bond length modulation and TI lattice distortion, and hence the coupling of electronic and lattice orders, could flatten the potential energy surface, resulting in a large shift in the chemical potential.

In contrast, the room temperature TCI phase does not exhibit such atomic bond length modulation and has a low DOS near the $E_F$,[26-27] which limits the transient states accessible in our relatively low laser fluence (0.6 mJ/cm$^2$) excitation regime. As noted above, the maximum laser fluence on this sample was limited by space charge effects – thus, it was not possible to directly laser-excite the TI-to-TCI phase transition. Furthermore, we did not observe photoexcited signal immediately after the pump pulse. We speculate that in the TCI phase, the ground state and excited states associated with the Cd – As substructure that we probe are optically dark to our EUV probe. Such optically dark states are determined by the relative phase between the sub-lattices and has recently been reported in a static ARPES study.[40] Future theoretical studies are needed to validate this hypothesis.

It is likely that multiple bond length modulations that are characteristic of the low-temperature TI phase[27] are excited as coherent phonon modes after laser excitation.[39, 41-42] However, real-time probing of these excitations would require techniques such as ultrafast X-ray scattering[43] or transient reflectivity[42]. Further dynamical probing with varying photon energy and higher electron energy resolution would be helpful to distinguish the bulk and surface state relaxation channels and to identify competing relaxation pathways. However, we note that the distorted lattice rearrangement in the TI phase will be independent of bulk state and surface states. Thus, as noted above, the measured slow rise time of the excited state and electron temperature could be a consequence of partial lattice relaxation of a distorted TI phase, which is known to be associated with the TI phase (see Fig. 1).[26] Small lattice rearrangements could introduce electronic structure changes near $E_F$ while leaving the more deeply bound electronic states intact.

In summary, we have explored the out-of-equilibrium states of NaCd$_4$As$_3$ in the topological crystalline insulator and topological insulator phases using tr-ARPES. A chemical potential shift of >150 meV was observed for the highest pump fluences in the TI phase after ultrafast laser excitation, which slowly rises and peaks after ~0.6 ps, and that persists for ~8 ps. The slow rise and fall time is likely related to the partial relaxation of the distorted lattice order in the TI phase after laser excitation, and an associated change in electronic order. In contrast, no distinct excited state is observed in the TCI phase after photoexcitation, which we attribute to the low density of states and phase space available near the Fermi level. Our results demonstrate how excitation by ultrafast light pulses can probe the excited states and interactions in phase-rich topological materials.

**Materials and Methods**

Material: The flux method was employed to grow single crystals of NaCd$_4$As$_3$. Na, Cd, and As were mixed in a ratio of 1:8:3 and placed in an alumina crucible, sealed in a quartz tube for heating. The details of the sample preparation and characterization are discussed elsewhere.[28]

Time- and angle-resolved photoemission spectroscopy (tr-ARPES): An amplified Ti:sapphire laser was used to generate the pump and probe pulses for this experiment, operating at a central wavelength of 786 nm (1.58 eV), with ~40 fs pulse duration, and 10 KHz repetition rate. The fundamental wavelength was frequency doubled to 393 nm, which was then used to generate the 7$^{th}$ order harmonic at a photon energy of 22.10 eV, via high harmonic generation (HHG) in Kr gas. This HHG pulse was then used to photoemit electrons after excitation of the material by the 786 nm pump pulse. The details of the HHG source and the time- and angle-resolved resolved photoelectron spectroscopy) setup can be found in our previous reports.[30-31, 44-46] The photoemitted electron intensity and kinetic energy at various angles were measured using a hemispherical electron analyzer, SPEC PHOIBOS 100. Prior to this measurement, the single crystal sample was cleaved in UHV vacuum better than $3\times10^{-10}$ Torr and measured at various temperatures, primarily at 78 K and 240 K, to access the TI and TCI phases of the material, respectively. The tr-ARPES spectra degraded over a period of ~5-8 hours for several samples measured - likely due to surface contamination from the presence of reactive Na atoms. To avoid any variation resulting from different samples and the limitation imposed by sample surface degradation over time, a controlled time-resolved measurement for a few representative fluences was done on the same sample to extract $\Delta T_e$. A similar procedure was applied to determine the chemical potential shift, but additional fluence measurements were made by measuring spectra at only two delay points – negative time delay and at 600 fs. The energy resolution is close to 130 meV, limited by the bandwidth of the ultrashort EUV pulses.

**Acknowledgement:**

M.M.M. and H.C.K. acknowledge support by the US Department of Energy, Office of Science, Basic Energy Sciences X-Ray Scattering Program Award DE-SC0002002 for this research. The ARPES setup was supported by the NSF through JILA Physics Frontiers Center PHY-2317149. J.K.F. was supported by the Department of Energy, Basic Energy Sciences under Award DE-FG02-08ER46542. J.K.F. was also supported by the McDevitt bequest at Georgetown University.

# Supplementary Information

# Non-equilibrium States and Interactions in the Topological Insulator and Topological Crystalline Insulator Phases of NaCd$_4$As$_3$

Tika R Kafle[1], Yingchao Zhang[1], Yi-yan Wang[2], Amy Lui[3], Xun Shi[1], Na Li[1], Richa Sapkota[1], Jeremy Thurston[1], Wenjing You[1], Shunye Gao[1], Qingxin Dong[2], Kai Rossnagel[3,4], Gen-Fu Chen[2], James K Freericks[5], Henry C Kapteyn[1,6], Margaret M Murnane[1]

[1]*Department of Physics and JILA, University of Colorado and NIST, Boulder, CO 80309, USA*
[2]*Institute of Physics and Beijing National Laboratory for Condensed Matter Physics*
[3]*Institute of Experimental and Applied Physics, Kiel University, D-24098 Kiel, Germany*
[4]*Ruprecht Haensel Laboratory, Deutsches Elektronen-Synchrotron DESY, D-22607 Hamburg, Germany*
[5]*Department of Physics, Georgetown University, Washington, DC 20057, USA*
[6]*KMLabs Inc., 4775 Walnut Street, #102, Boulder, Colorado 80301, USA*

* Corresponding author: tika.kafle@colorado.edu

## S1: Band structure characterization

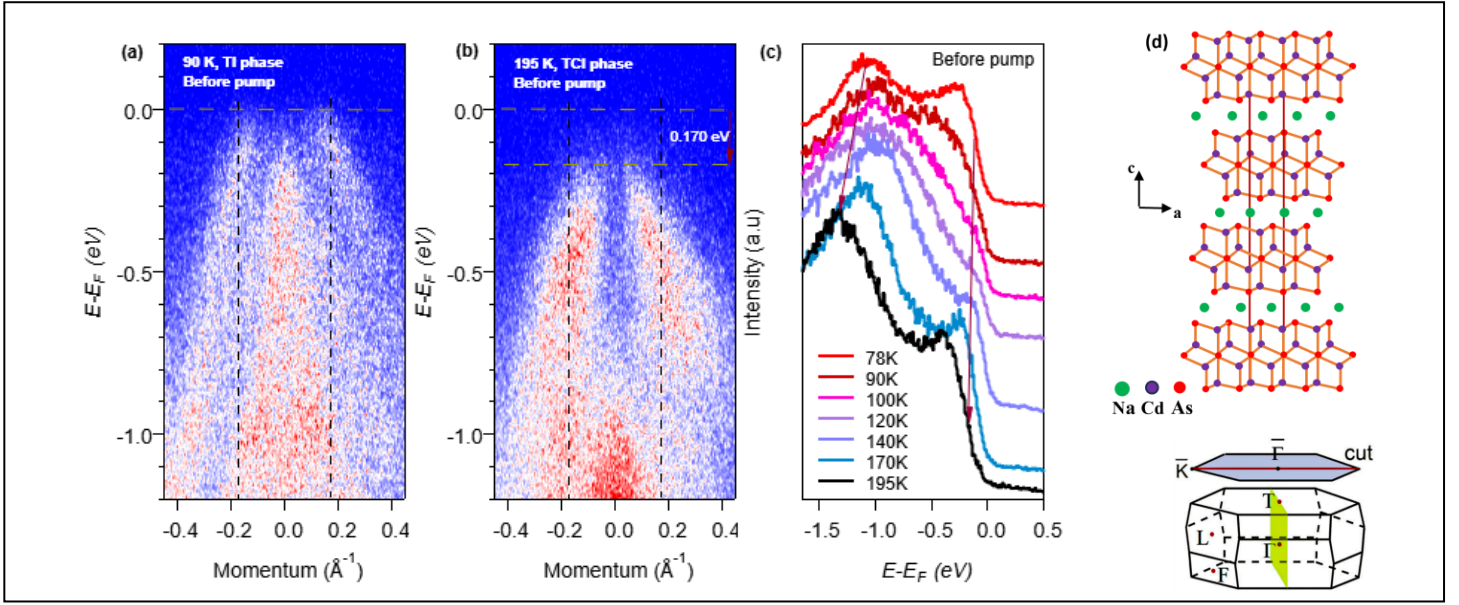

**Figure S1:** Equilibrium tr-ARPES spectra of (a) TI and (b) TCI phases of NaCd$_4$As$_3$ and the (c) EDCs taken at different temperatures, which agree with static ARPES data within our spectral resolution.[28] (d) The bottom figure is a schematic of the 3D Brillouin zone with high-symmetry points. The green shaded area represents the mirror plane of the room temperature crystal structure. The section at the top represents the direction along which the ARPES measurement is made. The top figure shows the full lattice structure of the TCI phase.

Here we characterize the material and identify two distinct topological phases. Figure S1 shows the equilibrium tr-ARPES data. In the low-temperature phase, there are linear-looking dispersive states on either side of a central band at the Γ point. The band dispersion, when compared with the previous results taken at a different photon energy (10 eV)[1], does not change significantly, so it is likely that the outer bands are surface states (SS). However, we note that multiple bands on either side of the Γ point, as reported in previous results, could not be resolved, mainly due to the limited energy resolution in our setup. In the room-temperature TCI phase (195 K and above), the entire band shifts down by ~0.15 eV towards higher binding energy. The EDCs, plotted around the Γ region near $E_F$ for different temperatures, show a distinct shift at 195 K. This change in electronic band structure validates the topological phase transition at ~195 K induced by structural changes. A thorough investigation with photon energy dependence and better energy resolution would be required for differentiating bulk bands and SS.

## S2: Determination of chemical potential shift and electron temperature

Here we discuss how we fit the momentum-integrated EDCs for all time delays to find the chemical potential shift and change in electron temperature shown in Fig. 3d of the main text. A Fermi-Dirac distribution multiplied by a Lorentzian function was used to fit the EDCs (see Eq. S1). A linear function was added to account for the background at different energy positions.

$$I(E) = (I_o + I_1 * E) + \left(\frac{A}{(E-E_{LP})^2 + \left(\frac{\Gamma}{2}\right)^2}\right)\left(\frac{1}{e^{\frac{E-E_F}{K_B T}} + 1}\right) \text{----------- (S1)}$$

The constants $I_0$ and $I_1$ are describe the background. The constant $A$ is the varying amplitude to account for spectral depletion/increment due to the pump laser pulse and any fluctuations in the laser. $E_{LP}$ is the Lorentzian peak center position, $\Gamma$ is the Lorentzian peak width, $E_F$ is the Fermi level (or chemical potential, μ), and $T$ corresponds to the electron temperature ($T_e$). For the accuracy and robustness of this method, we refer the reader to our previous work.[2-3] Note that the Lorentzian function in S1 is slightly modified to compensate for any change in the Lorentzian width within $A$. We find that the EDC peak position (marked by the vertical brown dashed line in Fig. S2a, $E_{LP}$) and the Lorentzian peak width just before the Fermi tail barely change for all time delays. Thus, $E_{LP}$ and $\Gamma$ are fixed, and a global fit is performed for all time-delay EDCs to determine $E_F$ and $T_e$. With the minimum number of fitting parameters, the fit of the Fermi edge looks reliable and accurate. The vertical blue dashed line is a guide to the eye to show the shift in $E_F$ position. The parameters $E_F$ and $T_e$ as obtained from the fit are shown in Fig. S2b.

Next, we used an alternative approach to quantify $E_F$ by determining the center of the Fermi edge using the built-in smoothing function of the Igor Pro software. The maximum and minimum intensities in the binding energy range between -0.5 eV and 0.5 eV were calculated and their average was taken to find the center position of the Fermi edge. The $E_F$ position obtained in this way (Fig. S2b) is in good agreement with that obtained from the Fermi-Dirac fit (Fig. S2b). Figure S2c shows the zoomed-in region as indicated by the dotted box in Fig. S2b. The peak $\Delta\mu$ shift appears at ~300 fs and persists for 1.5 ps. After 1.5 ps, it starts to relax to lower values. This is consistent with the EDCs, where the maximum leading band edge shift appears ~500 fs, persists for 1.5 ps, and starts to relax to lower values at 2 ps (Fig. S2d).

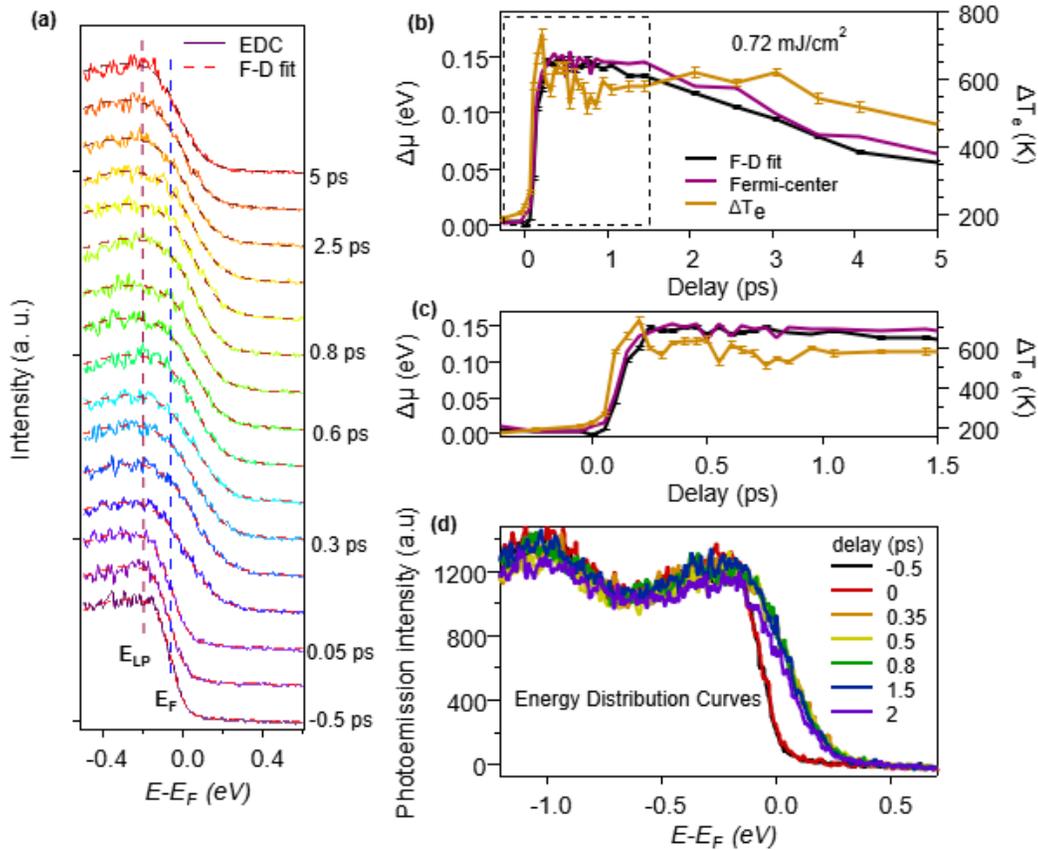

**Figure S2:** (a) Fermi-Dirac fit of the EDCs near $E_F$. The representative EDCs, colored lines, with corresponding delay times are shifted vertically and the fits are shown by dashed lines. The $\Delta\mu_F$ shift and $\Delta T_e$ obtained from the fit and the $E_F$ shift determined from the center of the Fermi edge are shown in (b) and a zoom-in region (dashed line) of the same plot is shown in (c). (d) The EDCs showing maximum $E_F$ shift around 500-800 fs, which persists for about 2 ps.

The $T_e$ obtained from the fit is normalized to the initial equilibrium lattice temperature. The rapid cooling of $T_e$ occurs within a ps, which is typical for electron-lattice thermalization. After a ps, the $T_e$ relaxation occurs by energy transfer to surface phonons which is a slower process. Note that there are signatures of oscillation in both electron temperature and Fermi level consistent with our assumption that phonon modes are coupled to electrons for the first few ps. A similar fitting procedure was used to determine $T_e$ and $\Delta E_F$ in the main text, Fig. 3, and other laser fluences.

**S3: Determination of the chemical potential shift in the TI and TCI phases and its dependence on the integration interval of momentum-integrated EDCs**

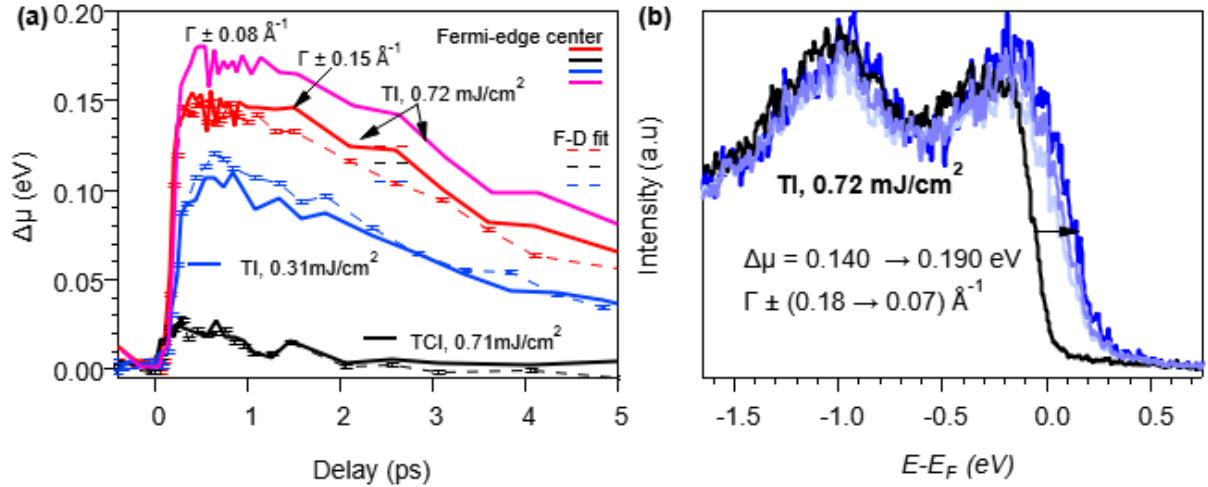

**Figure S3:** (a) $\Delta\mu_F$ as determined by the Fermi-edge center method, in the TI (90 K) and TCI (240 K) phases for different fluences and momentum integration intervals as indicated in the legend. The dotted line is the $\Delta\mu_F$ obtained from the Fermi-Dirac fit. The results of the two methods are in good agreement. (b) EDCs showing an increase in $\Delta\mu_F$ with decreasing momentum integration interval as discussed in main text Fig. 2.

## S4: Fluence dependence electron temperature rise time.

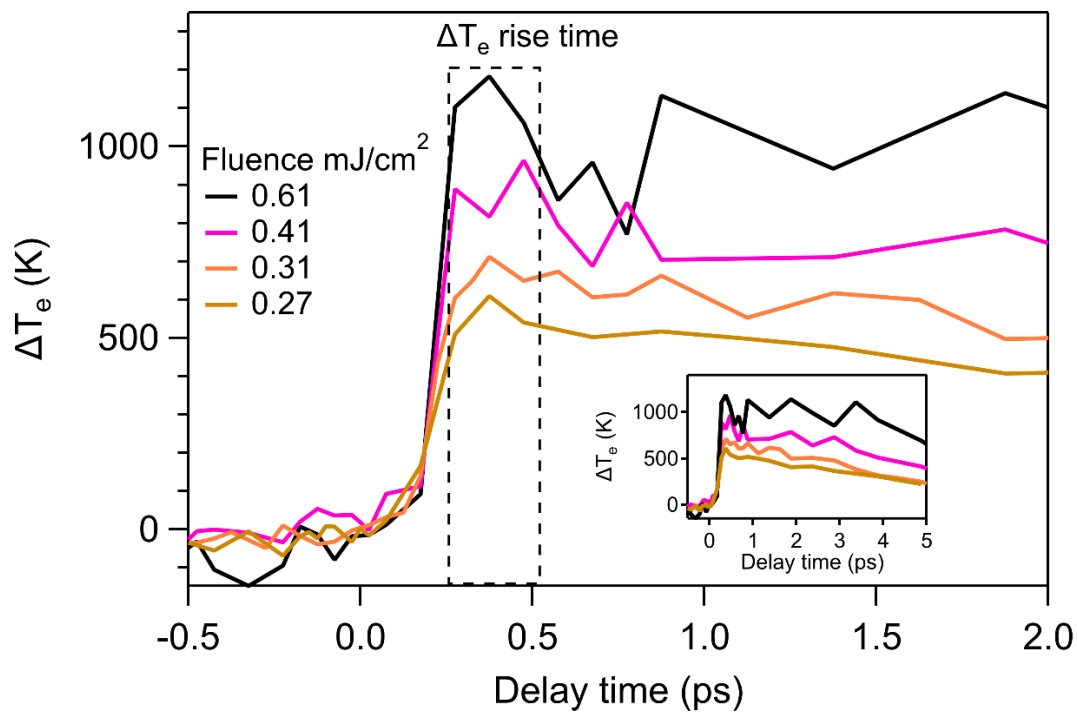

**Figure S4:** (a) Pump fluence-dependent electron temperature, $T_e$, rise time. The observed $T_e$ rise time of ~400 fs is independent of the pump fluence, but a gradual increase in electron temperature is observed, consistent with scattering due to higher electron density. The inset shows a slow decrease of $T_e$ after reaching its maximum.

## S5: Oscillatory signals in the excited state

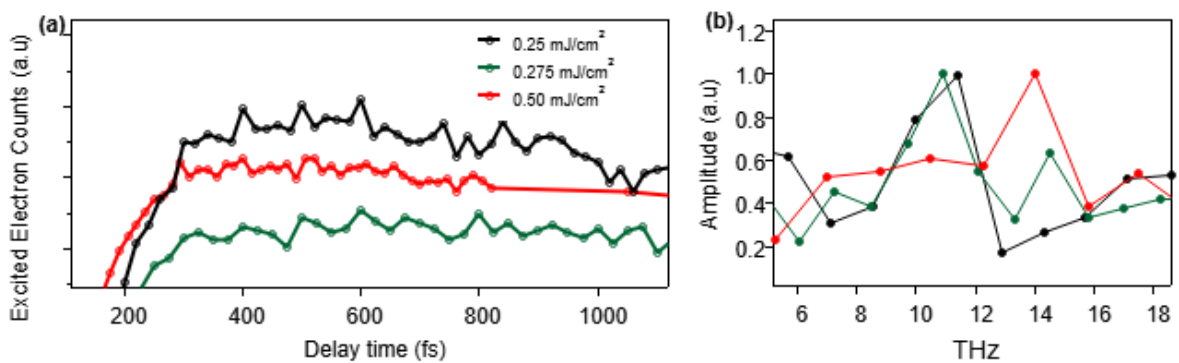

**Figure S5:** (a) The oscillatory signal of the excited state electrons at the pump fluence indicated in the legend and (b) the FFT of these signals. The measurement was performed with time steps of 15 fs (red), 20 fs (green) and 25 fs (black).